\documentclass[12pt]{article}
\usepackage{graphicx}
\usepackage{amssymb}
\usepackage{amsmath}
\newcommand{\be}{\begin{equation}}
\newcommand{\ee}{\end{equation}}
\newcommand{\ba}{\begin{eqnarray}}
\newcommand{\ea}{\end{eqnarray}}

\begin{document}

\begin{titlepage}
\begin{flushright}
\end{flushright}
\vfill
\begin{center}
{\Large\bf Leading Chiral Logarithms of $K_{S} \to \gamma \gamma$ and $K_{S} \to \gamma~\l^+ \l^-$  at two Loops}
\vfill
{\bf Karim Ghorbani\footnote{k-ghorbani@araku.ac.ir}  }\\[1cm]
{Physics Department, Faculty of Sciences, Arak University, Arak 38156-8-8349, Iran}
\end{center}
\vfill
\begin{abstract}
We obtain the leading divergences at two-loop order
for the decays $K_{S} \to \gamma \gamma$ and $K_{S} \to \gamma~\l^+ \l^-$
using only one-loop diagrams.
We then find the double chiral logarithmic corrections to 
the decay branching ratio of $K_{S} \to \gamma \gamma$ and to the decay 
rate for $K_{S} \to \gamma~\l^+ \l^-$.   
It turns out that these effects are numerically small and 
therefore make a very small enhancement on the branching ratio and decay rate.   
We also derive an expression for the corrections 
of type $\log \mu~\times$ LEC. Numerical analysis done 
for the process $K_{S} \to \gamma \gamma$ 
shows that these single logarithmic effects can be sizable but 
come with opposite signs with respect to the double chiral logarithms.

\end{abstract}
\vfill
keywords: Kaon rare decay; Weak and Strong Chiral Lagrangian; Leading Chiral Logarithm 
\vfill
{\footnotesize\noindent }

\end{titlepage}

\section{Introduction}
\label{int}
The non-leptonic kaon decay $K_{S} \to \gamma \gamma$
provides a decent testing bed for the effective chiral Lagrangian
method. The reason hinges in the fact that
to one-loop order, there is no short-distance effects due to the 
Furry's theorem \cite {Gaillard:1974hs} and the decay amplitude 
at one-loop order in chiral perturbation theory (ChPT) is 
free from unknown low energy constants (LECs)~\cite{DAmbrosio,Goity}.
For a good recent review on the kaon physics within the Standard Model 
one may consult \cite{Cirigliano:2011ny}.
The decay rate and branching ratio at one-loop order in ChPT
are evaluated in ~\cite{DAmbrosio,Goity}. It gives for the branching ratio
Br$(K_{S}\to \gamma \gamma) = 2.1\times 10^{-6}$.
This finding is in good agreement with the experimental measurement
of NA31 that obtained Br$(K_{S}\to \gamma \gamma) = (2.4 \pm 0.9)\times 10^{-6}$~\cite{NA31-1995}
and with that of KLOE that measured Br$(K_{S}\to \gamma \gamma) =
(2.26 \pm 0.12)\times 10^{-6}$~\cite{KLOE2007} .
On the other hand, the most recent measurement from NA48,
obtained Br$(K_{S}\to \gamma \gamma)= 2.71\times 10^{-6}$~\cite{NA48-2002}
with a total uncertainty of about $3\%$. 
The latter experiment opens up the possibility of a sizable
correction from effects beyond one-loop order.

It is demonstrated within the dispersion relation approach 
in \cite{Kambor1993tv,Truong:1993va} that in fact the $\pi~\pi$ rescattering
in the S-wave channel are important effects especially
for processes without counterterm contribution 
at one-loop order. They used in \cite{Kambor1993tv} 
the Pad\'e approximation for the 
Omn\`es function in the full unitarization procedure and 
found a significant enhancement in branching ratio,
Br$(K_{S}\to \gamma \gamma)= 2.3\times 10^{-6}$.
This brings the theoretical prediction into better agreement 
with the present experimental world average, 
Br$(K_{S}\to \gamma \gamma)= (2.63\pm 0.17) \times 10^{-6}$ \cite{PDG}. 

We now turn to the main point which motivates the present research.
What would be the size of the two-loop effects or order $p^6$ in chiral perturbation 
theory for the weak decay $K_{S} \to \gamma \gamma$? 
In fact, the presence of unknown low energy constants 
in the weak Lagrangian at next-to-leading order (NLO) have 
hindered our predictivity within ChPT at full two-loop order.         
However, of all different types of contributions to 
the full two-loop calculation, there is a part which is independent
of the unknown constants and can be evaluated using leading two-loop
divergences. These are the so-called Leading Logarithmic (LL) corrections.
Since the chiral LL corrections (single logs) are absent at one-loop order for
$K_{S} \to \gamma \gamma$, it is deemed interesting to investigate 
the importance of the LL contribution at two-loop order, i.e. chiral double log corrections.   
In the weak sector, the first study on the LL effects at two-loop 
order is done for the decay $K \to \pi \pi$ in \cite{Buchler:2005xn}. 
In the decay $K \to \pi \pi$, pion loop integral and kaon loop integral 
do not decouple in the subclass of two-loop diagrams
which are needed in order to find the double chiral log corrections.
So, for the process $K \to \pi \pi$, it was not possible 
to unambiguously define the double log corrections from only
one-loop calculations.    

Two comments are appropriate to mention. In the study of $\pi \pi$ scattering 
to two-loop accuracy within SU(2) ChPT, it was found that the bulk portion 
of the correction to the threshold parameters are due to the chiral 
logarithms \cite{Bijnens:1995yn}. This is not commonly the case in the 
SU(3) ChPT calculations. For instance, the double chiral logs 
in the vector form factor, $f_{+}(0)$, related to the semi-leptonic kaon decay
only make up about 25\% of the NNLO correction \cite{Bijnens:1998yu}. 
We are therefore curious to know what happens about the size of the 
leading logs for a process like $K_{S} \to \gamma \gamma$
with only two outgoing photons, reminding the fact that 
a combination of ChPT and dispersion relations can satisfactorily
predict the experimental data.  
       
Along the same line we study the decay $K_{S} \to \gamma~\l^+ \l^-$ with $l =e,\mu$.
Although this decay is not observed experimentally yet, 
it is interesting from the vantage point of having 
the same low energy structure as the decay $K_{S} \to \gamma \gamma$
to be investigated within ChPT.
       
A related rare process is $K_{S} \to \gamma^{\ast} \gamma^{\ast}$. 
In view of the recent LHCb measurement on the rare decay 
$K_{S} \to \mu^+  \mu^-$ \cite{LHCb:KSmumu}, other possible
rare decays at LHCb, namely, $K_{S} \to l^+ l^- l^+ l^-$ and 
$K_{S} \to l^+_{1} l^-_{1} l^+_{2} l^-_{2}$ \cite{LHCb} which occur
through the decay $K_{S} \to \gamma^{\ast} \gamma^{\ast}$ are studied 
in \cite{DAmbrosio:KSllll} emphasizing on the vector meson dominance contribution 
at ${\cal O}(p^6)$. These studies make the conclusion that at the prospect of 
the experimental data on the relevant $K_{S}$ and $K_{L}$ decays at LHCb, 
our theoretical predictions can be verified. It would be also interesting 
that with the LHCb upgrade we may get experimental data for the rare decay 
$K_{S} \to \gamma~\l^+ \l^-$.        
 
The structure of the article is as follows. A brief introduction to 
the weak and strong chiral Lagrangian up to NLO is given in Section~\ref{chpt}.
In Section~\ref{kin} the kinematics for the process $K_{S} \to \gamma \gamma$ 
and $K_{S} \to \gamma~\l^+ \l^- $ are discussed and one-loop result for
$K_{S} \to \gamma \gamma$ decay is reviewed in Section~\ref{oneloop}.   
The procedure in which we can derive the leading log corrections
and its link to the leading divergences are explained in Section~\ref{LLintro}.
Our analytical result concerning the single and double log correction 
are provided by ~Section~\ref{result}. Section~\ref{numerics} summarizes
our numerical results. The divergent part of the integrals are given in Appendix~A.

 \section{Chiral Lagrangians at ${\cal O}(p^2)$ and ${\cal O}(p^4)$} 
\label{chpt}
We employ chiral effective Lagrangians in order to study the low
energy dynamics of the strong and weak interactions.
The Lagrangians we use in the present work are the leading order 
and next-to-leading order chiral Lagrangians.
The expansion parameter is in terms of external momenta, $p$,
and quark masses, $m_{q}$.
Quark masses are counted of order $p^2$ due to the lowest
order mass relation $m_{\pi}^2 = B_{0}(m_{u}+m_{d})$.
Here we briefly discuss the leading order and next-to leading order 
strong and weak chiral Lagrangian.
The leading order Lagrangian which is of order $p^2$,
has the form
\be
{\cal L}_{2} = {\cal L}_{S2} + {\cal L}_{W2}.
\ee
The subscript in ${\cal L}_{2}$ indicates the chiral order.
${\cal L}_{S2}$ refers to the strong sector with $\Delta S = 0$
and ${\cal L}_{W2}$ stands for the effective weak interaction with $\Delta S = \pm 1$.
For the strong part we use~\cite{Weinberg:1968de}
\be
{\cal L}_{S2} = \frac{F_{0}^{2}}{4} \langle u_{\mu} u^{\mu}+ \chi_{+} \rangle,
\ee
where $F_{0}$ is the pion decay constant at chiral limit and we define the
matrices $u^{\mu}$ and $\chi_{\pm}$ as the following

\ba
u_{\mu} = i u^{\dag}D_{\mu}U u^{\dag} = u_{\mu}^{\dag}  \,, \quad u^{2} = U,
\nonumber\\
\chi_{\pm} = u^{\dag} \chi u^{\dag} \pm u\chi^{\dag}u.
\ea
The matrix $U \in SU(3)$ contains
the octet of light pseudo-scalar mesons with its exponential
representation given in terms of meson fields matrix as
\be
U(\phi) = \exp(i \sqrt{2} \phi/F_0)\,,
\ee
where
\ba
\phi (x)
 = \, \left( \begin{array}{ccc}
\displaystyle\frac{ \pi_3}{ \sqrt 2} \, + \, \frac{ \eta_8}{ \sqrt 6}
 & \pi^+ & K^+ \\
\pi^- &\displaystyle - \frac{\pi_3}{\sqrt 2} \, + \, \frac{ \eta_8}
{\sqrt 6}    & K^0 \\
K^- & \bar K^0 &\displaystyle - \frac{ 2 \, \eta_8}{\sqrt 6}
\end{array}  \right) .
\ea
We use the method of external fields discussed in~\cite{Gasser:1983yg}.
The external fields are then defined through the covariant derivatives as
\ba
D_{\mu} U = \partial_{\mu} U - i r_{\mu}U +iUl_{\mu}.
\ea
The right-handed and left-handed external fields are expressed
by $r_{\mu}$ and $l_{\mu}$ respectively. In the present work we set
\ba
r_{\mu} = l_{\mu}
 = e~A_{\mu} \left( \begin{array}{ccc}
\displaystyle 2/3 &   \\
    &\displaystyle -1/3 &  \\
 &   &\displaystyle -1/3
\end{array}  \right) .
\ea
The electron charge is denoted by $e$ and $A_{\mu}$ is the classical photon field.
The Hermitian $3\times3$
matrix $\chi$ involves the scalar (s) and pseudo-scalar (p) external
densities and is given by $\chi = 2B_{0}(s+ip)$.
The constant $B_{0}$ is related to the pion decay
constant and quark condensate. For our purpose it suffices to write
\ba
\chi
 = 2B_{0}\, \left( \begin{array}{ccc}
\displaystyle m_{u} &   \\
    &\displaystyle m_{d} &  \\
 &   &\displaystyle m_{s}
\end{array}  \right) .
\ea
The $\Delta S = \pm 1$ part of the weak effective Lagrangian contains both
the $\Delta I = 1/2$ piece and the $\Delta I = 3/2$ transition and has
the form~\cite{Cronin1967}
\ba
{\cal L}_{W2} =  F_{0}^{4} \Big[ G_{8} \langle \Delta_{32} u_{\mu} u^{\mu} \rangle
+G^{\prime}_{8} \langle \Delta_{32}\chi_{+} \rangle
\nonumber \\
+ G_{27} t^{ij,kl} \langle \Delta_{ij}u_{\mu}\rangle \langle \Delta_{kl}u^{\mu} \rangle \Big]
+ \text{h.c},
\ea
where the low energy constants $G_{8}$ and $G_{27}$ are defined in terms of  
dimensionless couplings $g_{8}$ and $g_{27}$ as
\ba 
G_{8,27} = -\frac{G_{F}}{\sqrt2} V_{ud} V^{*}_{us}~g_{8,27}. 
\ea

The matrix $\Delta_{ij}$ is given by
\ba
\Delta_{ij} = u \lambda_{ij} u^{\dag}  \,, \quad (\lambda_{ij})_{ab} = \delta_{ia}\delta_{jb}.
\ea
The non-zero components of the tensor $t^{ij,kl}$ are
\ba
t^{21,13} = t^{13,21} = \frac{1}{3} &\,,& \quad  t^{22,23} = t^{23,22} = -\frac{1}{6}
\nonumber\\
t^{23,33} = t^{33,23} = -\frac{1}{6} &\,,& \quad  t^{23,11} = t^{11,23} = \frac{1}{3}.
\ea
\begin{table}
\begin{center}
\begin{tabular}{|cc|cc|cc|}
\hline
$N_{i}$&  $n_{i}$         & $L_{i}$&  $l_{i}$&   $ H_{i}$&  $h_{i}$ \\
\hline
\hline
$N_{5}$   &  $3/2$       & $L_{1}$   & $3/32$  &   $ H_{1}$  &  $-1/8$  \\
$N_{7}$   &  $-9/8$       & $L_{2}$   & $3/16$  &   $ H_{2}$  &  $5/25$  \\
$N_{8}$   &  $-1/2$        & $L_{3}$   & $0$      &            &           \\
$N_{9}$   &  $3/4$       & $L_{4}$   & $1/8$  &            &           \\
$N_{10}$  &  $2/3$       & $L_{5}$   & $3/8$  &            &           \\
$N_{11}$  &  $-13/18$      & $L_{6}$   & $11/144$ &          &            \\
$N_{12}$  &  $-5/12$       & $L_{7}$   & $0$       &           &            \\
$N_{14}$  &  $1/4$       & $L_{8}$   & $5/48$   &           &            \\
$N_{15}$  &  $1/2$       & $L_{9}$   & $1/4$    &           &            \\ 
$N_{16}$  &  $-1/4$        & $L_{10}$   & $-1/4$    &           &            \\
$N_{17}$  &  $0$          &           &          &           &            \\
$N_{18}$  &  $-1/8$       &           &         &            &            \\
$N_{37}$  &  $-1/8$       &           &         &            &            \\
\hline
\end{tabular}
\end{center}
\caption{\label{LEC}The low energy constants of the strong and weak effective chiral Lagrangians at 
order $p^4$ which contribute to decays $K_{S} \to \gamma \gamma$ or $K_{S} \to \gamma~\l^+ \l^-$
are shown along with the coefficients of the divergent part.}
\end{table}
At order $p^4$, the chiral Lagrangian consists of two parts as 
\be
{\cal L}_{4} = {\cal L}_{S4} + {\cal L}_{W4}.
\ee
The SU(3) strong Lagrangian at next to leading order contains 10+2 
independent operators with corresponding low energy constants (LECs)\cite{Gasser:1984gg}
\begin{eqnarray}
\label{lagL4}
{\cal L}_{S4}&&\hspace{-0.5cm} =
L_1 \langle u_{\mu} u^{\mu} \rangle^2 
+L_{2} \langle u_{\mu} u^{\nu} \rangle \langle u^{\mu} u_{\nu} \rangle
+L_{3} \langle u_{\mu} u^{\mu} u_{\nu} u^{\nu} \rangle
+L_{4} \langle u_{\mu} u^{\mu} \rangle \langle \chi_{+} \rangle 
\nonumber\\&&\hspace{-0.1cm}
+L_{5} \langle u_{\mu} u^{\mu} \chi_{+} \rangle 
+L_{6} \langle \chi_{+} \rangle^2+ L_{7} \langle \chi_{-} \rangle^2  
+\frac{1}{2}L_{8} \langle \chi_{+}^2 + \chi_{-}^2 \rangle 
\nonumber\\&&\hspace{-0.1cm}
-iL_{9} \langle f^{\mu \nu}_{+} u_{\mu} u_{\nu} \rangle
+\frac{1}{4}L_{10} \langle f_{+\mu \nu} f_{+}^{\mu \nu} - f_{-\mu \nu} f_{-}^{\mu \nu}  \rangle
\nonumber\\&&\hspace{-0.1cm}
+\frac{1}{2}H_{1} \langle f_{+\mu \nu} f_{+}^{\mu \nu} + f_{-\mu \nu} f_{-}^{\mu \nu}  \rangle
+\frac{1}{4}H_{2} \langle \chi_{+}^2 - \chi_{-}^2 \rangle\,. 
\end{eqnarray}
Terms with $H_{1}$ and $H_{2}$ are only needed for renormalization and do not appear 
in physical processes. 
The field strength tensor is defined as 
\begin{eqnarray}
f_{\pm}^{\mu \nu}&&\hspace{-0.5cm} = u F^{\mu \nu}_{L} u^{\dag} \pm u^{\dag} F_{R}^{\mu \nu} u,
\nonumber\\
F^{\mu \nu}_{L}&&\hspace{-0.5cm} = \partial^{\mu}l^{\nu}-\partial^{\nu}l^{\mu}-i[l^{\mu},l^{\nu}], 
\nonumber\\
F^{\mu \nu}_{R}&&\hspace{-0.5cm} = \partial^{\mu}r^{\nu}-\partial^{\nu}r^{\mu}-i[r^{\mu},r^{\nu}]. 
\end{eqnarray}
In order to absorb the infinities arising from the loop integrals, the low 
energy constants need to be renormalized in an appropriate way. This is done
in \cite{Gasser:1984gg} by splitting the constants into a finite renormalized part 
and a infinite piece as 
\ba
L_{i}  =  L^{r}_{i} + \frac{l_{i}~\mu^{d-4}}{16\pi^2}~ \Big( \frac{1}{d-4} +c \Big)\,,
\nonumber \\
H_{i}  =  H^{r}_{i} + \frac{h_{i}~\mu^{d-4}}{16\pi^2}~ \Big( \frac{1}{d-4} +c \Big)\,,
\ea
where $d = 4-\epsilon$ is the space-time dimension in dimensional regularization. 
The constant $c$ depends on the regularization scheme used and for ChPT in the 
minimal subtraction scheme we have $c = -\frac{1}{2}(\log 4\pi +\gamma_{E}+1)$.
The coefficients $l_{i}$ and $h_{i}$ are listed in Table.~\ref{LEC}.  
   
The non-leptonic weak octet Lagrangian at NLO is discussed in full detail in \cite{Kambor1989tz}. 
The Lagrangian with all the terms relevant for the decays 
$K_S \to \gamma \gamma$ or $K_{S} \to \gamma~\l^+ \l^-$ is 
\begin{eqnarray}
\label{lagL4w}
{\cal L}_{W4}&&\hspace{-0.5cm} = F_{0}^2 G_{8} 
{\Big [ } N_{5} {\cal O}^{8}_{5} + N_{7} {\cal O}^{8}_{7} + N_{8} {\cal O}^{8}_{8} + N_{9} {\cal O}^{8}_{9} +
        N_{10} {\cal O}^{8}_{10} 
        \nonumber\\&&\hspace{-0.1cm}
        +N_{11} {\cal O}^{8}_{11}+ 
        N_{12} {\cal O}^{8}_{12} + N_{14} {\cal O}^{8}_{14} +
        N_{15} {\cal O}^{8}_{15} +N_{16} {\cal O}^{8}_{16} 
        \nonumber\\&&\hspace{-0.1cm}
        +N_{17} {\cal O}^{8}_{17} +N_{18} {\cal O}^{8}_{18} +
        N_{37} {\cal O}^{8}_{37} {\Big ]}. 
\end{eqnarray}
We use the basis for the operators in the Lagrangian above as set in \cite{Ecker:1992de}.
In the same fashion as we do in the strong Lagrangian, the weak LECs 
in the weak Lagrangian have to be renormalized in a proper way. 
The renormalization procedure is performed by evaluating the one-loop divergences 
in \cite{Kambor1989tz} with 
\begin{eqnarray}
N_{i}  =  N^{r}_{i} + \frac{n_{i}~\mu^{d-4}}{16\pi^2}~ 
\Big( \frac{1}{d-4} +c \Big)\,.
\end{eqnarray}
The constants $n_{i}$ are quoted in Table~\ref{LEC}.

\section{Kinematics for the decays $K_{S} \to \gamma \gamma$ and $K_{S} \to \gamma~\l^{+} \l^{-}$}
\label{kin}
The decay amplitude of $K_{S} \to \gamma \gamma$ with
the following momentum assignment
\ba
K_{S}(p)  \to   \gamma (q_{1}) \gamma (q_{2}),
\ea
has the form
\ba
A( K_{S} \to \gamma \gamma ) =  M_{\mu\nu}(q_{1},q_{2}) \hspace{.1cm}
{\epsilon_{1}}^{\mu} (q_{1}) \hspace{.1cm} {\epsilon_{2}}^{\nu}(q_{2}),
\ea
where ${\epsilon_{1}}^{\mu}$ and ${\epsilon_{2}}^{\nu}$ are the
polarization four-vectors of the outgoing photons
carrying momenta $q_{1}$ and $q_{2}$ respectively.
Due to the gauge invariance, Lorentz symmetry and Bose symmetry, $M_{\mu\nu}(q_{1},q_{2})$
takes on the specific form
\ba
M_{\mu\nu}(q_{1},q_{2}) = F(p^2)\hspace{.1cm}(  q_{1\nu} q_{2\mu} -
q_{1}.q_{2} \hspace{.1cm} g_{\mu\nu} ) \,,
\ea
where $p = q_{1}+q_{2}$ and $q_1^2 = q_2^2 = 0$ for photons with on-shell masses.
The decay width for a decay with two particles in the final state reads
\ba
\Gamma(K_{S} \to \gamma \gamma) = \frac{M_{K}^3}{64\pi} |F(p^2 = m_{K}^2)|^2.
\ea

The decay $K_{S} \to \gamma~\l^+ \l^- $ takes place via the decay $K_{S} \to \gamma \gamma^{\ast} $ 
with one photon being off-shell decaying into a lepton pair $e^+ e^-$ or $\mu^+ \mu^-$ .
The decay amplitude is parameterized as 
\ba
A( K_{S} \to \gamma~\l^{+} \l^{-} ) =  \frac{1}{q^2_{1}} M_{\mu\nu}(q_{1},q_{2}) \hspace{.1cm}
{\epsilon_{2}}^{\nu} (q_{2}) \hspace{.1cm}   {\bar u}(k) {\gamma}^{\mu} v(k^{\prime}),
\label{amplitudeKlepton}
\ea    
where, Lorentz gauge symmetry restricts $M_{\mu\nu}(q_{1},q_{2})$ to have the following form
\ba
M_{\mu\nu}(q_{1},q_{2}) = G(q^2_1)\hspace{.1cm}(  q_{1\nu} q_{2\mu} -
q_{1}.q_{2} \hspace{.1cm} g_{\mu\nu} ) \,.
\ea
The partial decay width for the process $K_{S} \to \gamma~\l^{+} \l^{-}$ normalized to the decay 
width of $K_{S} \to \gamma \gamma$ is 
\ba
\frac{1}{\Gamma_{K_{S} \to \gamma \gamma}} \frac{d\Gamma}{dz} = \frac{2}{z} (1-z)^3 \Big|\frac{G(z)}{G(0)}\Big|^2 
\frac{1}{\pi} \Im \Pi(z) \,, 
\ea
in which the electromagnetic spectral function related to the lepton pair is expressed by 
\ba
\frac{1}{\pi} \Im \Pi(z) = \frac{\alpha}{3\pi} \Big( 1+2\frac{r^2_{l}}{z}\Big)\sqrt{1-4r^2_{l}/z}~\Theta(z-4r^2_{l})\,,
\ea
where 
$r_{l} = m_{l}/m_{K}$ and $z = q^2_{1}/m^2_{K}$.

\section{ChPT result at ${\cal O}(p^4)$ for $K_{S} \to \gamma \gamma$ decay}
\label{oneloop}
\begin{figure}
\begin{center}
\includegraphics[width=.55\textwidth]{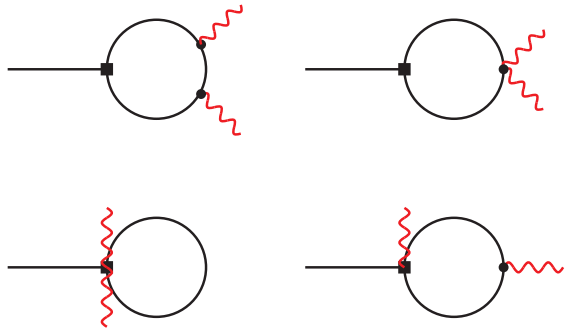}
\end{center}
\caption{Feynman diagrams of order $p^4$.
Solid lines represent the pseudo-scalar meson particles 
and wavy lines stand for photons.
{\tiny $\blacksquare$} is a $p^2$ weak vertex from ${\cal L}_{W2}$
and $\bullet$ is a $p^2$ strong vertex from ${\cal L}_{S2}$.
}
\label{loop4}
\end{figure}
The decay amplitude gets no tree-level contribution of order $p^{2}$ and $p^{4}$.
This is because all the particles involved here are neutral particles and 
on top of that due to the chiral symmetry.
Thus, the leading non-zero part of the amplitude originates 
from loop diagrams constructed
out of strong and weak Lagrangians of order $p^{2}$. The relevant Feynman diagrams
for this decay is depicted in Fig.~\ref{loop4}. Since tree diagrams are absent
here, we therefore expect that the sum of all the Feynman diagrams ends up
finite, i.e. all infinities from loop integrals vanish.
We show our result in a form that full agreement with the earlier formula 
given in \cite{DAmbrosio,Goity} can be simply understood. 
The following analytical result is achieved
\ba
\label{formula1}
F^{(4)}(p^2) &=& -\frac{2}{\pi} (G_8+\frac{2}{3}G_{27}) \alpha_{\text{em}} F_{0} \Big (\frac{p^2-m_{\pi}^2}{p^2} \Big)
\Big[ 1+\frac{m_{\pi}^2}{p^2}\log^2 \Big( \frac{\beta-1}{\beta+1} \Big) \Big]
\nonumber \\&&
-(m^{2}_{\pi} \to m^{2}_{K})\,,
\ea
where $\beta = \sqrt{1-4m_{\pi}^2/p^2}$. $m_{\pi}$ and $F_{0}$ are 
the bare pion mass and bare pion decay constant, respectively. 
One can see from the expression above that the pion loop contribution decouples from the kaon one.  
The bare parameters which appear in the decay amplitude makes
the definition of the amplitude at this order numerically ambiguous. 
One convenient way of resolving
the issue is to shift the bare quantities into their physical values but at the 
same time, one should keep track of all corrections which now go over to higher order.       
Hence, we define $F^{(4)}$ in terms of physical quantities such that 
$F^{(4)}$ = $F^{(4)}_{phys}$ + $F^{(6)}$. This can be done 
by correcting the bare parameters up to one-loop order as
$m^2_{\pi} = m^2_{\pi,phys} -\delta m^2$ for pion mass and $F_{0} = F_{\pi} - \delta F$
for the pion decay constant.
The corrections $\delta m^2$ and $\delta F$ provided by \cite{Gasser:1984gg} 
contain chiral logarithms and NLO low energy constants 
.
We present here only part of the correction $F^{(6)}$ 
which entails chiral logarithms and LECs: 
\ba
F^{(6)} = -\frac{2}{\pi} (G_8+\frac{2}{3}G_{27}) \frac{\alpha_{\text{em}}}{F_{0}}
        \Big\{-(1-\frac{m^2_{\pi}}{2p^2})L_{\pi} -\frac{1}{2} L_{K} + \frac{m^2_{\pi}}{6p^2} L_{\eta} 
 \nonumber\\&&\hspace{-10cm}
        -4(1+\frac{m^2_{\pi}}{p^2})\{ m^2_{\pi} L^{r}_{5}+(m^2_{\pi}+2m^2_{K}) L^{r}_{4} \}
        \nonumber\\&&\hspace{-10cm} 
       +16 \frac{m^2_{\pi}}{p^2} (m^2_{\pi}+2m^2_{K}) L^{r}_{6} 
       +16 \frac{m^4_{\pi}}{p^2}L^{r}_{8}
\Big\}\,,
\ea
where $L_{i} = \frac{m^2_{i}}{16\pi^2} \log(\frac{\mu^2}{m^2_{i}})$.
This type of contribution is necessary to be regarded when one considers 
the amplitude at full NNLO. The decay amplitude for related process 
$K_{S} \to \gamma~\l^+ \l^-$ at NLO is discussed in detail in \cite{Ecker:1987hd}.

\section{Leading Logarithms in ChPT}
\label{LLintro}
Chiral perturbation theory is a non-renormalizable field 
theory, in the sense that the cancellation of the infinities 
arising from loop integrals at a given order,
requires local operators with higher derivatives with 
respect to the lowest order Lagrangian.
In the strong sector for instance, there are only 
two operators in the leading Lagrangian, 10+2
operators in the next-to-leading Lagrangian and 90+4 
operators in the next-to-next-to-leading 
Lagrangian. The number of operators, thus, grows 
fast in going to higher orders. 
It was pointed out by Weinberg for the first time \cite{Weinberg:1978kz} 
that we can obtain information about the structure of 
the leading divergences at two-loop in a non-renormalizable 
field theory like ChPT by performing only one-loop calculations. 
In addition, this means that we can get the leading logarithmic 
corrections at two-loops from one-loop computations in ChPT. 
The generalization of this idea to any higher order is carried 
out in \cite{Buchler:2003vw}.
They derived in \cite{Buchler:2003vw} general relations 
that allows one to determine the leading
and subleading poles at any order in terms of one-loop diagrams.
In the following we recapitulate some results obtained in \cite{Bijnens:2009zi,Bijnens:2010xg} 
emphasizing on the relation which connect double chiral logarithmic 
corrections at two-loops to one-loop diagrams. 
Besides, we find that logarithmic corrections of type
$\log \mu~\times$~($L_{i}$ or $N_{i}$) can be obtained 
by determining singularities like ($L_{i}$ or $N_{i}$)/$\epsilon$.           

In general we can expand the bare Lagrangian with increasing power of $\hbar$ as
\ba
{\cal L} = {\cal L}_{(0)} + \hbar~{\cal L}_{(1)} + \hbar^2~{\cal L}_{(2)}+ ... \,, 
\ea
where ${\cal L}_{n}$ itself consists of a series of operators as
\ba
{\cal L}_{n} = \mu^{-n\epsilon} \sum_{i} c^{n}_{i} {\cal O}^{n}_{i}\,. 
\ea
The energy scale $\mu$ is defined such that the renormalized Lagrangian at all $\hbar$-order 
has space-time dimension $d$ where $d = 4-\epsilon$. 
The renormalization at a given order is achievable by subtracting the needed
infinities from the low energy constants in order to absorb the infinities coming from 
loop integrals and in the end to find finite result for a quantity at hand. 
Thus, it is necessary to write out the bare low energy 
constants $c^{n}_{i}$ as 
\ba
c^{n}_{i} =  c^{n}_{i,0} +\frac{1}{\epsilon} c^{n}_{i,1} + ... +\frac{1}{\epsilon^{n}} c^{n}_{i,n}\,. 
\ea   
We call $c^{n}_{i,0}$, the renormalized low energy constant to be determined from phenomenology.
Lets call $L^{n}_{l}$ loop diagrams of order $n$ with $l$ as the number 
of loops in the diagrams. A loop integral can be expanded in powers of poles in $\epsilon$, 
\ba
L^{n}_{l} = L^{n}_{l,0} + \frac{1}{\epsilon} L^{n}_{l,1}+...+\frac{1}{\epsilon^l} L^{n}_{l,l}\,. 
\ea   
We show first how renormalization procedure works at one-loop order. 
It is worth mentioning that it was proven in \cite{Buchler:2003vw} that the physical 
amplitude can be made finite with only taking into account the one-particle
irreducible diagrams at each order. 
At one-loop order, the loop diagrams are made out of vertices taken from lowest order 
Lagrangian, ${\cal L}_{(0)}$ and there is a contribution from counterterms 
taken from ${\cal L}_{(1)}$ which all together add up to 
\ba
\{c^{0}_{i}\} L^{1}_{1} + \mu^{-\epsilon} \{c^{1}_{i}\} L^{1}_{0}  = 
\{c^{0}_{i,0}\} L^{1}_{1,0}  + \frac{1}{\epsilon} \{c^{0}_{i,0}\} L^{1}_{1,1}
\nonumber \\&&\hspace{-6cm}
+\mu^{-\epsilon}\{c^{1}_{i,0}\}L^{1}_{0,0} + \frac{1}{\epsilon} \mu^{-\epsilon} \{c^{1}_{i,1}\} L^{1}_{0,0} \,,   
\ea      
where $\{...\}$ indicates the combination of all relevant low energy constants.
Taking into account the expansion $\mu^{-\epsilon} = 1-\epsilon \log \mu + ...$,
the cancellation of the infinities at one-loop order results in 
\ba
\{c^{1}_{i,1}\} L^{1}_{0,0}   = - \{c^{0}_{i,0}\} L^{1}_{1,1}\,.
\ea       
Applying the relation above, the $\log \mu$ dependent portion of the one-loop amplitude 
comes out  
\ba
-\{c^{1}_{i,1}\} L^{1}_{0,0}~ \log \mu = \{c^{0}_{i,0}\} L^{1}_{1,1}~ \log \mu. 
\ea 
At two-loop order the full expression contains both local and non-local divergences.  
The former contribution comes from a tree diagram derived from ${\cal L}_{(2)}$ and
one-loop diagrams with vertices from ${\cal L}_{(0)}$ and ${\cal L}_{(1)}$ as well as
two-loop diagrams with vertices from ${\cal L}_{(0)}$.  
The latter contribution originates from two-loop diagrams with vertices from Lagrangians ${\cal L}_{(0)}$ 
and from one-loop diagrams which involve vertices from both ${\cal L}_{(0)}$ and ${\cal L}_{(1)}$.   
We therefore can express the full result followed with expansion in $\epsilon$-poles as   
\ba
\mu^{-2\epsilon} \{c^{2}_{i}\}L^{2}_{0}+\mu^{-\epsilon} \{c^{1}_{i}\} L^{2}_{1}+ \{c^{0}_{i} \} L^{2}_{2} &=&
\mu^{-2\epsilon} \Big[ \{c^{2}_{i,0}\}+\frac{1}{\epsilon}\{c^{2}_{i,1}\}+\frac{1}{\epsilon^{2}}\{c^{2}_{i,2}\} \Big] L^{2}_{0,0}
\nonumber \\&&\hspace{-5cm}  
+\mu^{-\epsilon} \Big[ \{c^{1}_{i,0}\}+\frac{1}{\epsilon}\{c^{1}_{i,1}\} \Big] \Big[L^{2}_{1,0}+\frac{1}{\epsilon} L^{2}_{1,1} \Big]
\nonumber \\&&\hspace{-5cm}
+\{c^{0}_{i,0}\} \Big[ L^{2}_{2,0}+\frac{1}{\epsilon}L^{2}_{2,1}+\frac{1}{\epsilon^2}L^{2}_{2,2} \Big].
\label{2loop}
\ea  
We substitute the expansion $\mu^{-2\epsilon} = 1-2\epsilon \log \mu + 2\epsilon^2 \log^2 \mu+...$ 
into the 
above relation and ask for the cancellation of infinities of type $1/\epsilon^2$ 
and $\log \mu/\epsilon$, it turns out  
\ba
\{c^{2}_{i,2}\} L^{2}_{0,0} + \{c^{1}_{i,1}\} L^{2}_{1,1} + \{c^{0}_{i,0}\} L^{2}_{2,2} = 0 \,, 
\nonumber \\&&\hspace{-5.7cm}
2\{c^{2}_{i,2}\} L^{2}_{0,0} + \{c^{1}_{i,1}\}  L^{2}_{1,1} = 0\,. 
\ea 
The solution of the relations above reads
\ba
\{c^{0}_{i,0}\} L^{2}_{2,2} = -\frac{1}{2} \{c^{1}_{i,1}\} L^{2}_{1,1} \,,
\label{no1}
\ea
and 
\ba
\{c^{2}_{i,2}\} L^{2}_{0,0} = \{c^{0}_{i,0}\} L^{2}_{2,2} \,.
\label{no2}
\ea
We can now obtain the $\log^2 \mu$ dependent part of the full result at two-loop order by only 
collecting the relevant terms in Eq.~(\ref{2loop}) while we send $\epsilon \to 0$ and 
using Eq.~(\ref{no1}) and Eq.~(\ref{no2}) it gives   
\ba 
2~\{c^{2}_{i,2}\} L^{2}_{0,0}~\log^2 \mu+\frac{1}{2}~\{c^{1}_{i,1}\} L^{2}_{1,1}~\log^2 \mu  
= -\frac{1}{2}~\{c^{1}_{i,1}\} L^{2}_{1,1}~\log^2 \mu~.  
\label{final}
\ea
The final result found in Eq.~(\ref{final}) is important because it tells us 
that the coefficient of the leading 
logarithmic correction at two-loop order can be achieved by 
finding the double pole coefficient 
stemming from one-loop diagrams at next-to-next-to leading order.    

We are also interested to find corrections with single logarithms multiplied by 
the low energy constants. To this end, we turn back to Eq.~(\ref{2loop}) and restrict
our attention to terms with divergences as $1/\epsilon$. The cancellation of these
infinities requires the relation      
\ba 
\{c^{2}_{i,1}\} L^{2}_{0,0} + \{c^{1}_{i,0}\} L^{2}_{1,1} + \{c^{1}_{i,1}\} L^{2}_{1,0} +\{c^{0}_{i,0}\} L^{2}_{2,1}  = 0\,.   
\label{logeps}
\ea
Now we pick out terms proportional to $\log \mu$ in Eq.~(\ref{2loop}) 
and set $\epsilon \to 0$, the result is
\ba
-2 \{c^{2}_{i,1}\} L^{2}_{0,0}~\log \mu-\{c^{1}_{i,1}\} L^{2}_{1,0}~\log \mu 
   -\{c^{1}_{i,0}\} L^{2}_{1,1}~\log \mu  =
\nonumber \\&&\hspace{-10.5cm}
\{c^{1}_{i,1}\} L^{2}_{1,0}~\log \mu +\{c^{1}_{i,0}\} L^{2}_{1,1}~\log \mu
           +2\{c^{0}_{i,0}\} L^{2}_{2,1}~\log \mu \,,
\label{logNL}
\ea
where to get the equality we have used Eq.~(\ref{logeps}) in which the term
$\{c^{2}_{i,1}\} L^{2}_{0,0}$ is removed in favour of the rest.
Notice that the term $\{c^{1}_{i,0}\} L^{2}_{1,1}~\log \mu$ in the second line 
is considered to be 
the correction of type $L^{r}_{i}~\log \mu$ or $N^{r}_{i}~\log \mu$, in which 
$\{c^{1}_{i,0}\} L^{2}_{1,1}$ is the coefficient of the single $\epsilon$-pole.   
It should be noted that one may compute the single log corrections directly
using the one-loop diagrams but it sounds the easiest to follow the 
strategy discussed above.

\section{Calculation of the Leading Logarithm}
\label{result}
\begin{figure}
\begin{center}
\includegraphics[width=.85\textwidth]{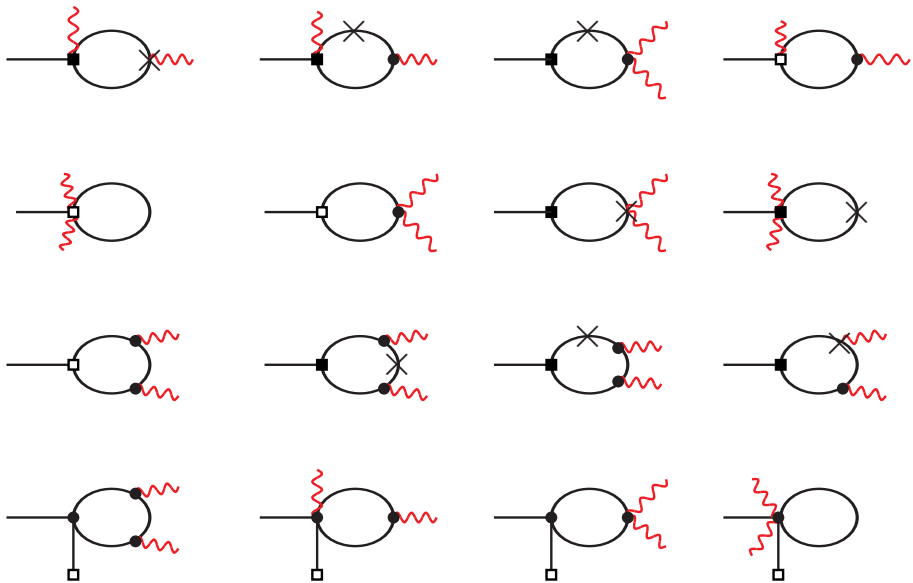}
\end{center}
\caption{A subset of Feynman diagrams of order $p^6$ which contribute to the 
double logarithms at two-loop order. The new vertices are:
{\tiny $\square$} is a $p^4$ weak vertex from ${\cal L}_{W4}$
and $\times$ is a $p^4$ strong vertex generated by ${\cal L}_{S4}$.}
\label{loop6}
\end{figure}
As we saw, at one-loop order there is no chiral logarithmic correction 
to the kaon decay to di-photon. 
Therefore, one expects the LL corrections to show up at two-loop order.   
We explained in the previous section that to obtain the double log corrections 
we only need to know the double poles from one-loop diagrams. 
All the necessary subdiagrams of order $p^6$ for the decays $K_{S} \to \gamma \gamma$ 
or $K_{S} \to \gamma~\l^+ \l^-$ are displayed in Fig.~\ref{loop6}. 
We only need the divergent part of the Feynman integrals so as to find the double pole
contribution of the full amplitude. 
In Appendix.~A we give the divergent part of the resulting integrals.     
We start with the process $K_{S} \to \gamma \gamma$.
Let us parameterize what we obtain here as ${\cal A}^{(6)}  = \{c^{1}_{i,1}\} L^{2}_{1,1}$, 
being the coefficient of the double pole divergences. Including
both pion and kaon loops the result explicitly is
\ba
{\cal A}^{(6)} = - \frac{4\pi \alpha_{\text{em}}}{(16\pi^2)^2}\frac{G_{8}}{F_{0}} \frac{1}{6} (m_{s}-\hat m)~(  q_{1\nu} q_{2\mu} -
q_{1}.q_{2} \hspace{.1cm} g_{\mu\nu} ) \,.
\ea   
This result to be gauge invariant is regarded as a non-trivial check on our analytical calculations. 
One additional way to verify the result is to note that Eq.~(\ref{no1}) restricts ${\cal A}^{(6)}$
to obey the relation ${\cal A}^{(6)} \propto {\cal A}^{(6)}_{\text{tree}}$, where ${\cal A}^{(6)}_{\text{tree}}$ is the amplitude
of the tree diagram of order $p^6$. The octet weak Lagrangian of order $p^6$ contains 
many operators but there is only one operator which can make the transition 
$K_{S} \to \gamma \gamma$ possible, see discussions in \cite{Buchalla:2003}.
The relevant Lagrangian is parameterized as
\ba
{\cal L}_{W6} = -i\frac{4\pi \alpha_{\text{em}}}{(16\pi^2)^2}~G_{8}~c_{1} F^{\mu\nu}F_{\mu\nu}  \langle \Delta \chi_{+} \rangle \,,   
\ea    
where $c_{1}$ is an unknown low energy constant.
It is then a straightforward task to find the decay amplitude as 
\ba
{\cal A}^{(6)}_{\text{tree}} = - \frac{4\pi \alpha_{\text{em}}}{(16\pi^2)^2}\frac{G_{8}}{F_{0}} 
\frac{8}{9}c_{1} (m_{s}-\hat m)~(  q_{1\nu} q_{2\mu} - q_{1}.q_{2} \hspace{.1cm} g_{\mu\nu} ) \,.
\ea 
We are therefore convinced that the relation ${\cal A}^{(6)} \propto {\cal A}^{(6)}_{\text{tree}}$ holds
and ${\cal A}^{(6)}$ has the correct structure.      
One important observation which turns out from our direct computation 
is that the pion loop integrals can be disentangled from the kaon 
loop integrals in our expression\footnote{This is not the case however, when 
we look at the full set of Feynman diagrams at two-loop order.}. 
Our result with only pion integrals reads
\ba
{\cal A}^{(6)}_{\pi} = \frac{4\pi \alpha_{\text{em}}}{(16\pi^2)^2}\frac{G_{8}}{F_{0}} \frac{1}{3} \hat m~( q_{1\nu} q_{2\mu} -
q_{1}.q_{2} \hspace{.1cm} g_{\mu\nu} ) 
- \frac{4\pi \alpha_{\text{em}}}{(16\pi^2)^2}\frac{G_{8}}{F_{0}} \frac{2}{9} (q_{1}.q_{2}) q_{1\mu} q_{2\nu}\,,
\ea
and with only kaon integrals leads to 
\ba
{\cal A}^{(6)}_{K} = - \frac{4\pi \alpha_{\text{em}}}{(16\pi^2)^2}\frac{G_{8}}{F_{0}} \frac{1}{6} (m_{s}+\hat m)~( q_{1\nu} q_{2\mu} -
q_{1}.q_{2} \hspace{.1cm} g_{\mu\nu} ) 
\nonumber \\&&\hspace{-8.6cm}
+ \frac{4\pi \alpha_{\text{em}}}{(16\pi^2)^2}\frac{G_{8}}{F_{0}} \frac{2}{9} (q_{1}.q_{2}) q_{1\mu} q_{2\nu}\,.
\ea  
In the two relations above, terms proportional to $q_{1\mu} q_{2\nu}$ do not contribute to the
physical amplitude since photons in the process here are on the mass-shell and consequently
$q_{1}.\epsilon_{1} = q_{2}.\epsilon_{2} = 0$.
With the formula provided by Eq.~(\ref{final}) we are now able to write down our formula for 
the leading log correction
\ba 
F^{(6)}_{\text{LL}} = -\frac{4\pi \alpha_{\text{em}}}{(16\pi^2)^2}\frac{G_{8}}{F_{0}} \frac{1}{24} 
\Big(m^2_{\pi} \log^2(\frac{m^2_{\pi}}{\mu^2})-m^2_{K} \log^2(\frac{m^2_{K}}{\mu^2}) \Big)\,,
\ea
where employed are the leading order mass relations, 
$m^2_{\pi} = 2B_{0} \hat m$ and $m^2_{K} =B_{0} (m_{s}+\hat m)$. 

In addition, it is of interest to find analytically logarithmic corrections 
of type $\log \mu \times L^{r}_{i}$ and $\log \mu \times N^{r}_{i}$. It is explained 
in section.~\ref{LLintro} that we only need to sequester divergent terms
as $L^{r}_{i}/\epsilon$ and $N^{r}_{i}/\epsilon$ and then 
with the application of Eq.~(\ref{logNL}) we find 
\ba
F^{(6)}_{\log \times \text{LEC}} = -\frac{4\pi \alpha_{\text{em}}}{32\pi^2}\frac{G_{8}}{F_{0}}
 \Big\{ \frac{16}{3}m^{2}_{\pi}(N^{r}_{37} +2N^{r}_{18}+N^{r}_{15}-N^{r}_{14})
\nonumber \\&&\hspace{-9cm}    
  +(L^{r}_{9}+L^{r}_{10})(16~p^2-32~m^{2}_{\pi}) \Big\}~\log(\frac{m_{\pi}^2}{\mu^2})
     - (m^{2}_{\pi} \to m^{2}_{K}).
\ea
We redo our calculations for the process $K_{S} \to \gamma~\l^{+} \l^{-}$. 
In this case we should keep in mind that $q^2_{2}$ = 0 but $q^2_{1} \neq 0$.
The coefficient of the double pole divergences including both pion and kaon 
integrals gives rise to 
\ba
{\cal B}^{(6)} = - \frac{4\pi \alpha_{\text{em}}}{(16\pi^2)^2}\frac{G_{8}}{F_{0}} \frac{1}{6} (m_{s}-\hat m)~(  q_{1\nu} q_{2\mu} -
q_{1}.q_{2} \hspace{.1cm} g_{\mu\nu} ) \,.
\ea   
This is identical to ${\cal A}^{(6)}$, the amplitude of $K_{S} \to \gamma \gamma$.
It is also possible for this process to separate the contribution of the pion 
integrals and kaon integrals.
Taking only pion integrals into account we obtain for the amplitude
\ba
{\cal B}^{(6)}_{\pi} = \frac{4\pi \alpha_{\text{em}}}{(16\pi^2)^2}\frac{G_{8}}{F_{0}} \Big(
\frac{1}{3} \hat m~( q_{1\nu} q_{2\mu} - q_{1}.q_{2} \hspace{.1cm} g_{\mu\nu} ) 
- \frac{2}{9} (q_{1}.q_{2}) q_{1\mu} q_{2\nu} -\frac{1}{9} q^2_{1} q_{2\mu} q_{2\nu} \Big) \,,
\ea
and taking only kaon integrals we find 
\ba
{\cal B}^{(6)}_{K} = - \frac{4\pi \alpha_{\text{em}}}{(16\pi^2)^2}\frac{G_{8}}{F_{0}} 
\Big( \frac{1}{6} (m_{s}+\hat m)~( q_{1\nu} q_{2\mu} - q_{1}.q_{2} \hspace{.1cm} g_{\mu\nu} ) 
\nonumber \\&&\hspace{-8.6cm}
+ \frac{2}{9} (q_{1}.q_{2}) q_{1\mu} q_{2\nu} + \frac{1}{9} q^2_{1} q_{2\mu} q_{2\nu}  \Big) \,.
\ea  
When we put these results back into Eq.~(\ref{amplitudeKlepton}) 
terms proportional to $q_{2\nu}$ vanish because for the on-shell photon 
we have $q_{2}.\epsilon_{2} = 0$. 
We make use of the formula in Eq.~(\ref{final}) and obtain the 
leading log effects for the decay $K_{S} \to \gamma \l^{+} \l^{-}$ as
\ba 
G^{(6)}_{\text{LL}} = -\frac{4\pi \alpha_{\text{em}}}{(16\pi^2)^2}\frac{G_{8}}{F_{0}} \frac{1}{24} 
\Big(m^2_{\pi} \log^2(\frac{m^2_{\pi}}{\mu^2})-m^2_{K} \log^2(\frac{m^2_{K}}{\mu^2}) \Big)\,.
\ea   
The bottom line here is the observation that the leading log corrections for the decays 
$K_{S} \to \gamma \gamma$ and $K_{S} \to \gamma~\l^{+} \l^{-}$ are identical.  
We finish this subsection by presenting our calculations 
on the contributions of type $\log \mu \times$ ($L^r_{i}$ or $N^r_{i}$) for 
the decay $K_{S} \to \gamma~\l^{+} \l^{-}$. Our results read
\ba
G^{(6)}_{\log \times \text{LEC}} = -\frac{4\pi \alpha_{\text{em}}}{32\pi^2}\frac{G_{8}}{F_{0}}
 \Big\{ \frac{16}{3}m^{2}_{\pi}(N^{r}_{37} +2N^{r}_{18}+N^{r}_{15}-N^{r}_{14})
\nonumber \\&&\hspace{-9cm}    
-\frac{2}{3}q^{2}_{1}(N^{r}_{17} +N^{r}_{16}+N^{r}_{15}+N^{r}_{14})
\nonumber \\&&\hspace{-9cm} 
  +(L^{r}_{9}+L^{r}_{10})(32~q_{1}.q_{2}-32~m^{2}_{\pi}+16~q^2_{1}) \Big\}~\log(\frac{m_{\pi}^2}{\mu^2})
\nonumber \\&&\hspace{-9cm} 
     - (m^{2}_{\pi} \to m^{2}_{K}).
\ea

\section{Numerical results}       
\label{numerics}

\begin{table}
\begin{center}
\begin{tabular}{c|ccc}
\hline
$Br(K_{S}\to \gamma \gamma)$  &  NLO & NLO+LL &  NLO+LL+          \\
    $\times 10^6$             &      &        &  (log$\times$LEC)  \\
\hline
$\mu$ = 0.50 GeV             &2.0399& 2.0407 &   1.827         \\
$\mu$ = 0.77 GeV             &2.0399& 2.0401 &   1.877         \\
$\mu$ = 1.00 GeV             &2.0399& 2.0387 &   1.912         \\

\hline
\end{tabular}
\end{center}
\caption{\label{numeric1} The branching ratios for the decay $K_{S} \to \gamma \gamma$ at 
three different values of the renormalization scale are compared including 
the double chiral log corrections.}
\end{table}

\begin{table}
\begin{center}
\begin{tabular}{c|cccccc}
\hline
 LEC $\times 10^{3}$  & $L^{r}_{9}$ & $L^{r}_{10}$ & $N^{r}_{14}$ & $N^{r}_{15}$ & $N^{r}_{18}$ & $N^{r}_{37}$   \\
\hline
$\mu = 0.5$~GeV  & 6.61 & -4.74 & -9.71 & 7.31 & -0.34 & -0.34   \\
$\mu = 0.77$~GeV  & 5.93 & -4.06 & -10.4 & 5.95 & 0 & 0          \\
$\mu = 1.0$~GeV  & 5.51 & -3.64 & -10.8 & 5.12 & 0.20 & 0.20   \\
\hline
\end{tabular}
\end{center}
\caption{\label{LECs} Shown are the renormalized low energy constants
at three different values of the renormalization scale $\mu$.
The weak constants $N^{r}_{14}$ and $N^{r}_{15}$ are given in \cite{Bijnens:2004vz}
at $\mu = 0.77$ GeV and we choose $N^{r}_{18} = N^{r}_{37} = 0$.}
\end{table}

We are now ready to evaluate numerically the leading log contribution 
to the decay branching ratio. Let us begin with the NLO amplitude for the 
$K_{S} \to \gamma \gamma$ decay given in Eq.~(\ref{formula1}).
As input we use for the masses $m_{\pi^+}$ = 0.136~GeV and $m_{K}$ = 0.497~GeV
and for the pion decay constant $F_{\pi}$ = 0.0924~GeV.
There is an ambiguity in knowing which values for 
the weak coupling $g_{8}$ and $g_{27}$ should be used
at this level of calculations. 
It is found  $g_{8} = 4.99$ and $g_{27} = 0.297$ 
from a fit to the decay $K \to \pi \pi$ at leading order,
see discussions in \cite{Cirigliano:2011ny}.
At NLO fit, $g_{8}$ receives a rather significant reduction 
such that $g_{8} = 3.62$ and $g_{27} =0.286$ \cite{Cirigliano:2011ny}. 
Since in this research, it is the matter of comparing the size of the LL effects 
with NLO result, it may suffice to use the leading order values 
of the weak couplings, namely, $g_{8} = 4.99$ and $g_{27} =0.297$. 
We also use the same values as introduced above when we compute
the LL effects.  
In Table.~\ref{numeric1} we compare the decay branching ratio of 
$K_{S} \to \gamma \gamma$ both at one-loop order and with the 
inclusion of the LL effects 
at three different values of the renormalization 
scale, namely, $\mu$ = 1~GeV, 0.77~GeV and 0.5~GeV.
As it is evident, the double log correction is the largest at $\mu$ = 0.5~GeV
and changes very little with varying $\mu$. But at any rate, the size of 
the correction is meager even though it goes in the right direction.

Moreover, we estimate the size of the single log effects.
The values for the LECs used in the numerical calculations 
are listed in Table.~\ref{LECs}.
Our numerical results shown in Table.~\ref{numeric1} indicate
that these effects are significantly larger in magnitude 
than the LL corrections as expected, but they go in the opposite direction 
with respect to the LL effects.

Finally we calculate numerically the decay width of $K_{S} \to \gamma~\l^+ \l^-$
normalized to the decay width of $K_{S} \to \gamma \gamma$ both 
for electron pair and muon pair in the final state. 
Input parameters are the same as those we used for the 
decay $K_{S} \to \gamma \gamma$. 
Our calculations summarized in Table.~\ref{numeric2} compare the NLO results 
and the NLO+LL effects at different renormalization scales. It is seen that 
the LL contribution has a very small impact on the decay width, though 
having the largest contribution at $\mu = 0.5$ GeV.

\begin{table}
\begin{center}
\begin{tabular}{c|cccc}
\hline
$\frac{\Gamma(K_{S}\to \gamma \l^+ \l^-)}{\Gamma(K_{S}\to \gamma \gamma)}$  &  NLO  &  NLO+LL &  NLO      & NLO+LL     \\
                                                                          & $l=e$ &  $l= e$ & $l = \mu$ & $l=\mu$    \\
\hline
$\mu$ = 0.50 GeV      & 1.5967$\times10^{-2}$& 1.5969$\times10^{-2}$ & 3.6924$\times10^{-4}$ & 3.6930$\times10^{-4}$    \\
$\mu$ = 0.77 GeV      & 1.5967$\times10^{-2}$& 1.5967$\times10^{-2}$ & 3.6924$\times10^{-4}$ & 3.6926$\times10^{-4}$    \\
$\mu$ = 1.00 GeV      & 1.5967$\times10^{-2}$& 1.5964$\times10^{-2}$ & 3.6924$\times10^{-4}$ & 3.6915$\times10^{-4}$    \\
\hline
\end{tabular}
\end{center}
\caption{\label{numeric2} The decay width of the process $K_{S}\to \gamma~\l^+ \l^-$
normalized to the decay width of $K_{S} \to \gamma \gamma$ at 
three different values of the renormalization scale are compared including 
the double chiral log corrections.}
\end{table}

\section{Conclusions}
\label{con}
The calculation of the leading logarithmic corrections at two-loop order 
has been the main aim behind the present work. 
These effects are the only part of the NNLO result that can be obtained
from one-loop calculations and are free from unknown constants.
We knew already from earlier works that LL effects
are the sub-dominant part of the NNLO. In the case of $K \to \gamma \gamma$, earlier 
findings based on dispersion relation technique suggests that 
the LL correction might be even smaller than those found in other studied processes.
We have shown numerically that the size of the leading log corrections is very small indeed.
Relying on earlier experiences and our finding here, we can confirm  
that the full NNLO correction cannot enhance the branching ratio significantly.  
In addition, we found analytically the single log corrections of
type $\log \mu \times$ LEC as part of the higher order effects. 
It turned out that these corrections are numerically large 
but we know that these will go through a cancellation among 
different contributions in the full NNLO.     

We have also studied the LL effects in the decay width of the processes
$K_{S} \to \gamma~e^+ e^-$ and $K_{S} \to \gamma~\mu^+ \mu^-$.
It turned out that double log corrections are also very small
in these decay channels .    
   
\section{Acknowledgments}
\label{Ack}
I would like to thank Johan Bijnens and Lund University
for financial support and warm hospitality during my visit 
in the Institute of High Energy Physics. I am also very grateful
to Johan Bijnens for helpful discussions. The FORM code \cite{Form3}
is used in our analytical calculations.

\section{Appendix~A: One-loop integrals}
\label{Apen}
As we mentioned in the text, for our purpose  we only need  the divergent piece 
of the resulting one-loop integrals.   
For integrals with more than one propagator we need to make use of the 
Feynman parameter formula and then we should redefine the momentum variable
to get an integral with powers of one propagator. We only present here our final result.
\ba
\frac{1}{i} \int \frac{d^dq}{(2\pi)^d} \frac{1}{q^2-m^2} = 
\frac{m^2}{16\pi^2}\frac{2}{4-d}+ \text{finite}\,,
\ea

\ba
\frac{1}{i} \int \frac{d^dq}{(2\pi)^d} \frac{1}{(q^2-m^2)^2} = 
\frac{1}{16\pi^2}\frac{2}{4-d}+ \text{finite}\,,
\ea
\ba
\frac{1}{i} \int \frac{d^dq}{(2\pi)^d} \frac{1}{(q^2-m^2)((q+p)^2-m^2)} = 
\frac{1}{16\pi^2}\frac{2}{4-d}+ \text{finite}\,,
\ea

\ba
 \frac{1}{i} \int \frac{d^dq}{(2\pi)^d}
\frac{q^{\mu}}{(q^2-m^2)((q+p)^2-m^2)} = 
\frac{p^{\mu}}{16\pi^2}\frac{1}{(4-d)}+ \text{finite}\,,
\ea

\ba
\frac{1}{i} \int \frac{d^dq}{(2\pi)^d}
\frac{q^{\mu} q^{\nu}}{(q^2-m^2)((q+p)^2-m^2)} =  
 \frac{2}{{16\pi^2}(4-d)}(\frac{p^{\mu} p^{\nu}}{3}+g^{\mu\nu})+\text{finite}\,,
\ea 
\ba
\frac{1}{i} \int \frac{d^dq}{(2\pi)^d}
\frac{q^{\mu} q^{\nu}}{(q^2-m^2)((q+p)^2-m^2)((q+p+s)^2-m^2)}  
\nonumber \\&&\hspace{-7cm}  
= \frac{2}{{16\pi^2}(4-d)}\frac{g^{\mu\nu}}{16}+\text{finite}\,,
\ea 

\ba
\frac{1}{i} \int \frac{d^dq}{(2\pi)^d}
\frac{q^{\mu} q^{\nu} q^{\alpha}}{(q^2-m^2)((q+p)^2-m^2)((q+p+s)^2-m^2)}  
\nonumber \\&&\hspace{-8cm}  
= \frac{2}{{16\pi^2}(4-d)}
  \Big( \frac{p^\mu g^{\nu\alpha}+p^\nu g^{\mu\alpha}+p^\alpha g^{\mu\nu}}{6} \Big)
\nonumber \\&&\hspace{-7.6cm}
  +\frac{2}{{16\pi^2}(4-d)} 
  \Big( \frac{s^\mu g^{\nu\alpha}+s^\nu g^{\mu\alpha}+s^\alpha g^{\mu\nu}}{12} \Big)+\text{finite}\,,
\ea 
where $d = 4 -\epsilon$. We do not show integrals which are finite.

\end{document}